\documentclass[aps,prb,amsmath,amssymb,twocolumn,longbibliography, superscriptaddress]{revtex4-2}
\usepackage{graphicx}
\usepackage{dcolumn}
\usepackage{bm}
\usepackage{mathrsfs}
\usepackage{subfigure}
\usepackage{natbib}
\usepackage{amssymb}
\usepackage{multirow}
\usepackage{float}
\usepackage{color}
\usepackage{ulem}
\usepackage{amsmath}
\usepackage[colorlinks,linkcolor=blue,anchorcolor=blue,citecolor=blue,urlcolor=blue]{hyperref}

\begin{document}

\title{Universal crossover in surface superconductivity: Impact of varying Debye energy}

\author{Quanyong Zhu}
\affiliation{School of Mathematics and Computer Science, Lishui University, 323000 Lishui, China}

\author{Xiaobin Luo}
\affiliation{Zhejiang Key Laboratory of Quantum State Control and Optical Field Manipulation, Department of Physics, Zhejiang Sci-Tech University, 310018 Hangzhou, China}

\author{A. A. Shanenko}
\affiliation{Moscow Center for Advanced Studies, Kulakova str. 20, Moscow 123592, Russia}
\affiliation{HSE University, 101000 Moscow, Russia}

\author{Yajiang Chen}
\email{yjchen@zstu.edu.cn}
\affiliation{Zhejiang Key Laboratory of Quantum State Control and Optical Field Manipulation, Department of Physics, Zhejiang Sci-Tech University, 310018 Hangzhou, China}

\date{\today}
\begin{abstract}
Recently, interference-induced surface superconductivity (SC) has been predicted within an attractive Hubbard model with $s$-wave pairing, prompting intensive studies of its properties. The most notable finding is that the surface critical temperature $T_{cs}$ can be significantly enhanced relative to the bulk critical temperature $T_{cb}$. In this work, considering a $1D$ attractive Hubbard model for the half-filling level, we investigate how this enhancement is affected by variations in the Debye energy $\hbar\omega_D$, which controls the number of states contributing to the pair potential and, in turn, influences the critical temperature. Our study reveals a universal crossover of the surface SC from the weak- to strong-coupling regime, regardless of the specific value of the Debye energy. The location of this crossover is marked by the maximum of $\tau = (T_{cs} - T_{cb})/T_{cb}$, which depends strongly on $\hbar\omega_D$. At its maximum, $\tau$ can increase up to nearly $70\%$. Additionally, we examine the evolution of the ratio $\Delta_{s0}/k_B T_{cs}$ along the crossover, where $\Delta_{s0}$ is the zero-temperature pair potential near the surface (the chain ends), and demonstrate that this ratio can significantly deviate from $\Delta_{b0}/k_B T_{cb}$, where $\Delta_{b0}$ is the zero-temperature bulk pair potential (in the chain center). Our findings may offer valuable insights into the search for higher critical temperatures in narrow-band superconductors.  
\end{abstract}
\maketitle

\section{Introduction}

The physics of surface superconductivity (SC) has been a major focus in the fields of superconductivity and functional materials since the pioneering theoretical work by Saint-James and de Gennes~\cite{saint-james1963}. Their work demonstrated that the surface SC persists at the magnetic fields between the second and third critical fields, a finding later confirmed in metallic alloys, such as Pb$_x$Tl$_{1-x}$~\cite{bonmardion1964,tomasch1964}, PbIn~\cite{gygax1964}, PbBi~\cite{strongin1964a} and NbTa~\cite{hempstead1964a}. At a certain field, the surface critical temperature $T_{cs}$, which is higher than the bulk critical temperature $T_{cb}$, is needed to destroy the surface SC. To characterize the surface enhancement, it is convenient to use the parameter $\tau = (T_{cs} - T_{cb})/T_{cb}$.

Additionally, other mechanisms of the surface SC enhancement not involving a magnetic field have been proposed, following Ginzburg's seminal work~\cite{ginzburg1964}. These predictions have been extensively investigated in both theory and experiments, including the surface SC induced by specific surface phonon modes~\cite{ginzburg1964, strongin1968,dickey1968,naugle1973,chen2012b}, surface electron states~\cite{zhao2020a, nagai2016}, and superconducting proximity effects~\cite{mcmillan1968a, zhitomirsky2001, buzdin2005}. 

Recently, a novel mechanism for the surface SC enhancement, induced by the near-surface constructive interference of itinerant quasiparticles in the absence of a magnetic field, was reported within the framework of the attractive Hubbard model with the nearest-neighbor hopping and $s$-wave pairing~\cite{croitoru2020, chen2022}. (It is important to stress that the effect of interest does not arise within the parabolic band approximation~\cite{giamarchi1990, croitoru2020}.) Its several nontrivial properties have been investigated. Notably, it was shown that $T_{cs}$ can be tailored by a weak external static electric field, while $T_{cb}$ remains unchanged~\cite{bai2023a}. It was also demonstrated that this surface SC exhibits a Fulde-Ferrel-Larkin-Ovchinnikov state in external magnetic fields~\cite{barkman2019}, shows significant enhancements at the sample corners in $2D$ and $3D$ cases~\cite{samoilenka2020a}, manifests a complex competition between itinerant quasiparticles and topological bound states in the proximitized topological insulators~\cite{chen2024c, wang2023d}, and persists in multiband superconductors~\cite{benfenati2021}. 

It was also found that the surface SC is sensitive to the interplay between the pair coupling strength $g$ and the chemical potential $\mu$. In particular, it was shown~\cite{samoilenka2020a} that the relative surface SC enhancement $\tau$, taken as a function of $g$, exhibits a maximum dependent on $\mu$. It was proved~\cite{samoilenka2020} that this feature appears in $1D$, $2D$, and $3D$ variants of the attractive Hubbard model, except for the situation with the chemical potential close to the band edges in the $1D$ case. The maximal value of $\tau$ is most pronounced in the $1D$ case, where $\tau$ was found to rise up to $25\%$ for half-filling~\cite{samoilenka2020}. In the $2D$ and $3D$ cases, additionally to the surface SC, there are corner SC enhancements, where $\tau$ can go up to $40\%$, see Fig. 7 in Ref.~\onlinecite{samoilenka2020}.  

However, the study in Ref.~\onlinecite{samoilenka2020} was limited to the case where the Debye energy $\hbar\omega_D$ is larger than half the single-particle bandwidth, and did not explore the combined effects of varying both $g$ and $\hbar\omega_D$ in the favored regime of half-filling. Reducing the Debye energy below half the bandwidth, modifies the number of quasiparticle states contributing to the pair potential, which can significantly influence the near-surface interference of the pair states. For instance, it was reported~\cite{bai2023} that when $\hbar\omega_D$ goes below half the bandwidth, $\tau$ can increase to nearly $70\%$, which is significantly larger than the results obtained in Ref.~\onlinecite{samoilenka2020}. Therefore, further investigations are required. 

In the present work, we investigate how the interference-induced surface SC depends on both the coupling strength $g$ and the Debye energy $\hbar\omega_D$ for half-filling. For simplicity, we focus on a $1D$ attractive Hubbard model with $s$-wave pairing. Our analysis is based on numerically solving the self-consistent Bogoliubov-de Gennes (BdG) equations, from which we extract the critical temperatures $T_{cs}$ and $T_{cb}$, and then calculate $\tau$. Our results reveal the presence of a universal crossover in the surface SC from the weak- to strong-coupling regimes, regardless of a particular Debye energy. This crossover is accompanied by a pronounced maximum of $\tau$~(up to nearly $70\%$), the value and position of which depend strongly on $\hbar\omega_D$. The corresponding ratio $\Delta_{s0}/k_BT_{cs}$~(where $\Delta_{s0}$ is the zero-temperature pair potential at the chain ends) deviates significantly from both its bulk counterpart and textbook BCS value. Throughout the paper, we use the term ``surface" to refer to the behavior near the chain ends, maintaining a connection to higher dimensions, as our qualitative results on the surface SC are general and applicable to the $2D$ and $3D$ cases.

The paper is organized as follows. The BdG equations for a $1D$ attractive Hubbard model with $s$-wave pairing are outlined in Sec.~\ref{II}. Section ~\ref{III} presents our results and discussions. Finally, concluding remarks are provided in Sec.~\ref{IV}.

\section{Theoretical Formalism}\label{II}

As is mentioned in the Introduction, our consideration is based on a $1D$ attractive Hubbard model with $s-$wave pairing and nearest-neighboring hopping controlled by the amplitude $t$~\cite{croitoru2020,tanaka2000}. The model Hamiltonian reads 
\begin{equation}\label{H}
H = -\sum_{i\sigma,\eta=\pm1} t\,c^\dagger_{i+\eta,\sigma} c_{i\sigma} - \sum_{i\sigma}\mu c^\dagger_{i\sigma} c_{i\sigma} -g\sum_i n_{i\uparrow}n_{i\downarrow},
\end{equation}
where $\mu$ and $g$ are the chemical potential and the pair coupling strength, respectively. $c_{i\sigma}$ and $c^\dagger_{i\sigma}$ are the annihilation and creation operators of an electron at the site $i$ ($=1,$..., $N$) with spin projection $\sigma$ ($=\uparrow,\, \downarrow$), where $N$ is the total site number. Additionally, $n_{i\sigma}$ is the site-dependent electron number operator. By applying the mean-field approximation~\cite{gennes1966}, Eq.~(\ref{H}) is reduced to the effective mean-field Hamiltonian of the form 
\begin{equation}\label{Heff}
    H_{\rm eff}  = \sum_{ij\sigma}H_{ij}c_{i\sigma}^\dagger c_{j\sigma} + \sum_i \big[\Delta(i)c_{i\uparrow}^\dagger c_{i\downarrow}^\dagger+ \Delta^*(i)c_{i\downarrow} c_{i\uparrow}\big],
\end{equation}
where $H_{ij} = -t (\delta_{i,j-1} + \delta_{i,j+1}) - \mu\delta_{ij}$, with $\delta_{ij}$ being the Kronecker delta function, and $\Delta(i)$ is the site-dependent superconducting pair potential. Notice that the Hartree-Fock potential is ignored in Eq.~(\ref{Heff}) as it merely introduces an extra confinement potential~\cite{chen2009,chen2024} and does not qualitatively change our conclusions. 

By diagonalizing $H_{\rm eff}$ with the Bogoliubov-Valatin transformation~\cite{ketterson1999}, the following BdG equations are obtained 
\begin{subequations}\label{bdg}
\begin{align}
 \sum_j H_{ij} u_\alpha(j) + \Delta(i) v_\alpha(i) &= \varepsilon_\alpha u_\alpha(i), \\
\Delta^*(i) u_\alpha(i) - \sum_j H^*_{ij} v_\alpha(j) &= \varepsilon_\alpha v_\alpha(i).
\end{align}
\end{subequations}
Here, $\varepsilon_\alpha$, $u_\alpha(i)$ and $v_\alpha(i)$ are the quasiparticle energy and wavefunctions, respectively. The self-consistent condition for $\Delta(i)$ is derived by comparing the free energies for the effective and full Hamiltonians, which results in 
\begin{equation}\label{Delta}
\Delta(i) = g \sum_{0\le\varepsilon_\alpha\le\hbar\omega_D} u_\alpha(i)v^*_\alpha(i)[1-2f(\varepsilon_\alpha)],
\end{equation}
with $f(\varepsilon_{\alpha})$ being the Fermi-Dirac distribution function. The summation range in Eq.~(\ref{Delta}) indicates that only the physical branch of the quasiparticle spectrum should be taken into account~\cite{gennes1966}. Higher-energy quasiparticles contribute more significantly to the surface SC compared to lower-energy ones (strictly speaking in the absence of external electric fields, see Fig. 3 in Ref.~\cite{bai2023}). This explains why the value of $\hbar\omega_D$ is so important in studying the surface SC. 

In addition, when solving the BdG equations, one should take into consideration that the electron density $n_e(i)$ is also connected to the quasiparticle specifications by the relation 
\begin{equation}\label{ne}
    n_e(i) = 2\sum_\alpha f(\varepsilon_\alpha)\,|u_\alpha(i)|^2 + [(1-f(\varepsilon_\alpha)]\,|v_\alpha(i)|^2.
\end{equation} 
This relation determines the chemical potential $\mu$ for a given electron filling level $\bar{n}_e=(1/N)\sum_i n_e(i)$. 

\begin{figure}[t]
\centering
\includegraphics[width=1\linewidth]{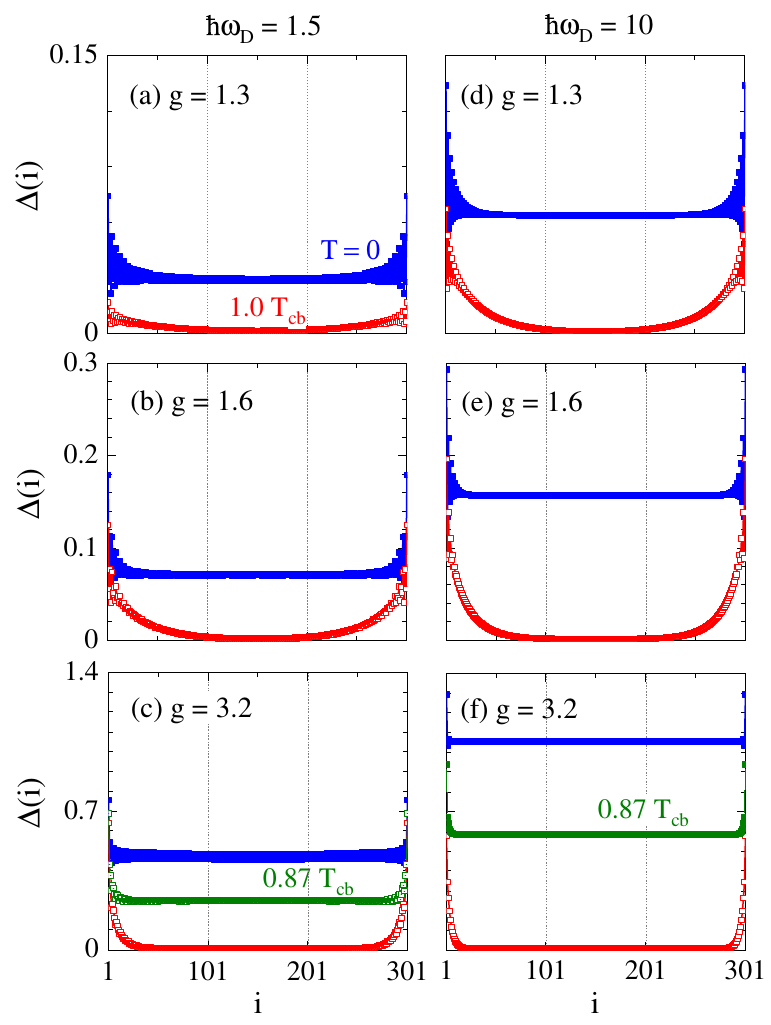}
\caption{The pair potential $\Delta(i)$ versus the site index $i$, as calculated for $g=1.3$, $1.6$, $3.2$, with $\hbar\omega_D=1.5$ (a-c), $10$ (d-f). The blue curves represent data for $T=0$, while the green and red ones are for $T=0.87\,T_{cb}$ and $1.0\,T_{cb}$, respectively.}
\label{fig1}
\end{figure}

To self-consistently solve the BdG equations~(\ref{bdg}), we begin by applying initial trial values for the pair potential $\Delta(i)$ and the chemical potential $\mu$. These initial values are used to construct the matrix eigenvalue equation from the BdG equations given by Eq.~(\ref{bdg}). Solving this eigenvalue problem provides the corresponding quasiparticle spectrum and wavefunctions, from which updated values for $\Delta(i)$ and the electron density $n_e(i)$ are calculated according to Eqs.(\ref{Delta}) and (\ref{ne}). Given a fixed electron filling level $\bar{n}_e$, a new $\mu$ can then be determined. The process continues iteratively, with the updated $\Delta(i)$ and $\mu$ being reinserted into the BdG equations, until both quantities converge according to the desired level of accuracy.

In our calculations, the hopping parameter $t$ can be scaled out so that $\Delta$, $\mu$, $\varepsilon_\alpha$, $g$, $\hbar\omega_D$, and the band width are measured in units of $t$. The temperature (along with $T_{cs}$ and $T_{cb}$) is given in units of $t/k_B$, where $k_B$ is the Boltzmann constant. The number of sites is set to $N=301$, which is large enough to avoid quantum-size effects~\cite{chen2009,chen2012}. Finally, the accuracy of the self-consistent calculation of $\Delta(i)$ and $\mu$ is set to $10^{-8}$.

\section{Results and discussions}\label{III}

To illustrate the influence of the coupling strength $g$ and Debye energy $\hbar\omega_D$ on the surface SC, Fig.~\ref{fig1} shows profiles of the pair potential $\Delta(i)$ calculated for $g=1.3$, $1.6$, and $3.2$ with $\hbar\omega_D=1.5$ and $10$ for several temperatures. The blue, green, and red curves represent data for $T/T_{cb}=0$, $0.87$, and $1.0$, respectively. For the bulk critical temperature, we have $T_{cb}=0.020$, $0.044$ and $0.290$ for $g=1.3$, $1.6$, and $3.2$ at $\hbar\omega_D=1.5$, while $T_{cb} = 0.041$, $0.095$, and $0.599$ at $\hbar\omega_D=10$. Here we deliberately choose the two typical scenarios: one is for the Debye energy larger than half the bandwidth $2$, while the other is for $\hbar\omega_D$ smaller than $2$. The last scenario means that all the physical branch quasiparticles are included in the summation of Eq.~(\ref{Delta}), while the first scenario implies significant limitations on the higher-energy quasiparticles. 

At $T=0$, all the pair potentials shown in Figs.~\ref{fig1}(a-f) are enhanced near the chain ends ($i=1$ and $301$), which, in higher dimensions, corresponds to the surface SC. For both $\hbar\omega_D=1.5$ and $10$ at $T=0$, the larger the value of $g$, the greater the pair potential near the boundaries (the surface pair potential). For the same value of $g$, the surface pair potential for $\hbar\omega_D=10$ is higher than that of $\hbar\omega_D=1.5$, due to a larger number of the contributing quasiparticles in the case of $\hbar\omega_D=10$. For all relevant parameters, the bulk pair potential (in the chain center) vanishes at $T/T_{cb}=1.0$, while the surface pair potential remains finite at higher temperatures $T_{cb}< T < T_{cs}$. Figure~\ref{fig1} shows that the region with the zero pair potential around the center of the system increases as $g$ rises. Consequently, the spatial extent (and the related characteristic length) over which the condensate near the boundaries persists decreases with increasing $g$.  

\begin{figure}[t]
\centering
\includegraphics[width=1\linewidth]{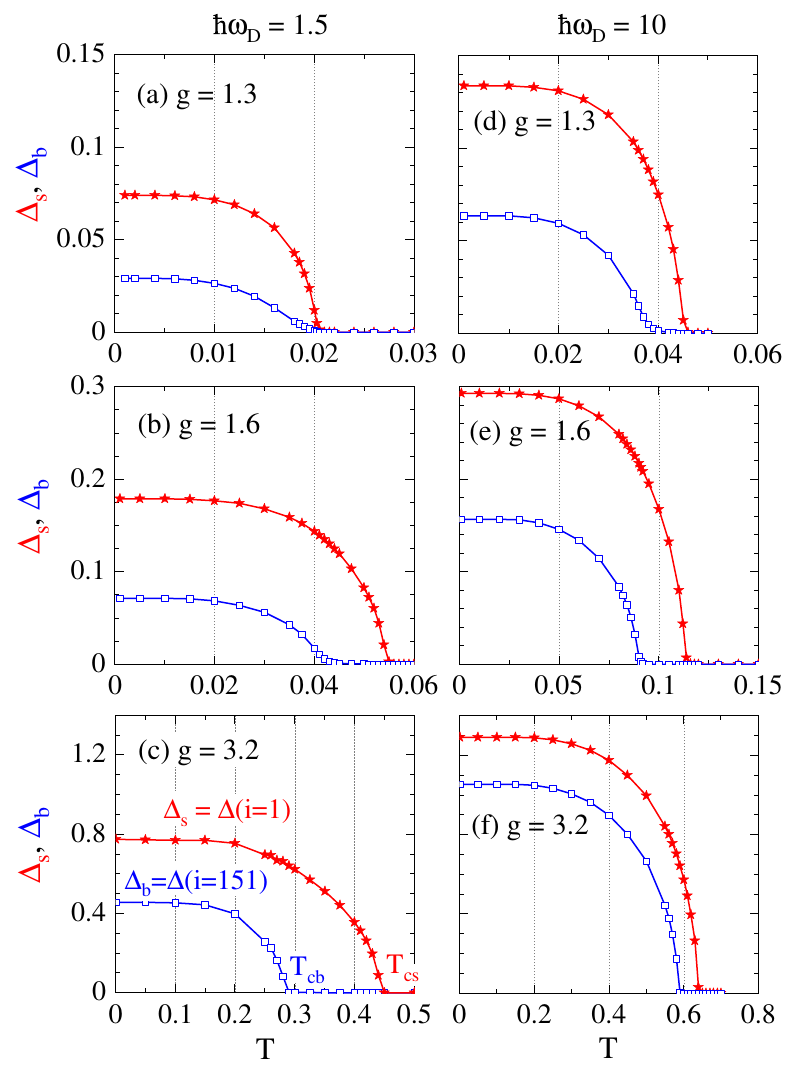}
\caption{The surface (at the chain ends; curves with red stars) and bulk (in the chain center; curves with blue squares) local pair potentials, $\Delta_s = \Delta(i=1,,301)$ and $\Delta_b = \Delta(i=151)$, are shown as functions of temperature $T$ for $g=1.3$, $1.6$, and $3.2$, with $\hbar\omega_D=1.5$ (a-c) and $\hbar\omega_D=10$ (d-f). The surface and bulk critical temperatures ($T_{cs}$ and $T_{cb}$), where $\Delta_s$ and $\Delta_b$ respectively drop to zero, are highlighted in panel (c).}
\label{fig2}
\end{figure}

In Fig.~\ref{fig2}, the temperature-dependent surface and bulk pair potentials, $\Delta_s = \Delta(i=1) = \Delta(i=N)$ and $\Delta_b = \Delta(i=151)$, are presented for $g=1.3$, $1.6$, and $3.2$ at $\hbar\omega_D=1.5$~(a-c) and $10$~(d-f). As $T$ increases, both $\Delta_s(T)$ and $\Delta_b(T)$ decrease, eventually approaching zero at $T_{cs}$ and $T_{cb}$, respectively. For $\hbar\omega_D=1.5$, we find $T_{cs}=0.0210$, $0.0545$, and $0.445$ at $g=1.3$, $1.6$, and $3.2$, respectively. The corresponding values of the surface SC enhancement parameter are calculated as $\tau=5.0\%$, $23.9\%$, and $53.4\%$. In this case, the higher the value of $g$, the greater the value of $\tau$. For $\hbar\omega_D=10$, we find $T_{cs} = 0.0455$, $0.115$, and $0.645$ at $g=1.3$, $1.6$, and $3.2$, with the corresponding surface SC enhancements of $\tau=13.8\%$, $26.4\%$, and $9.3\%$, respectively. Unlike the case of $\hbar\omega_D=1.5$, $\tau$ does not increase with $g$ for $\hbar\omega_D=10$. Thus, we conclude that variations in $\hbar\omega_D$ significantly affect the $g$-dependence of the surface SC enhancement parameter $\tau$. 

\begin{figure}[t]
\centering
\includegraphics[width=0.95\linewidth]{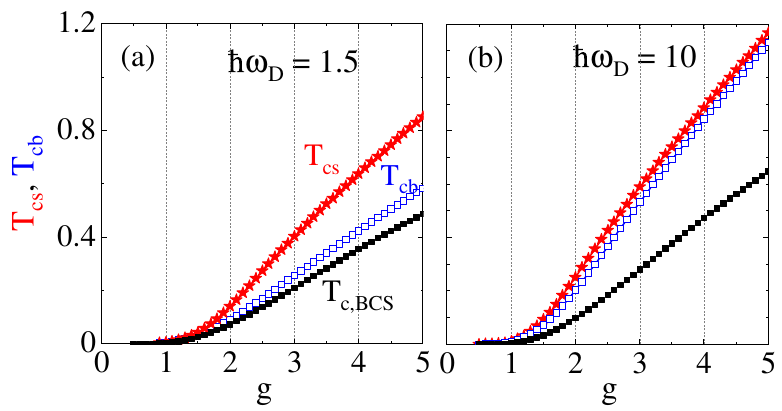}
\caption{$T_{cs}$ and $T_{cb}$ as functions of the coupling strength $g$ for $\hbar\omega_D=1.5$ (a) and $10$ (b). The results of $T_{cs}$ and $T_{cb}$ are given by red-starred and blue-squared curves, respectively. The black curves with solid squares represent the BCS critical temperature $T_{c,{\rm BCS}}$.}
\label{fig3}
\end{figure}

To delve deeper, Fig.~\ref{fig3} illustrates the surface and bulk critical temperatures as functions of the coupling strength $g$ for $\hbar\omega_D = 1.5$ and $10$. The red-starred curves represent the data for $T_{cs}$, while the blue-squared curves correspond to $T_{cb}$. Additionally, the black curves with solid squares depict the BCS critical temperature $k_BT_{c,\,{\rm BCS}}=1.134 \hbar\omega_D{\rm e}^{-1/gD(\mu)}$, with $D(\mu)$ being the normal single-electron density of states (DOS)~\cite{gennes1966}. $D(\mu)$ is extracted from the single-electron spectrum of the tight-binding model $\xi_n=-2t{\rm cos}(k_na)$, with $k_na=\pi n / (N+1)$, $n=1,\,...,\,N$ and $a$ being the lattice constant~\cite{tanaka2000}. At $\mu=0$, corresponding to half-filling, we have $k_na=\pi/2$ and $D(\mu) = \frac{N+1}{N-1}\cdot \frac{1}{2\pi {\rm sin}(k_na)} \approx 1/2\pi$~(DOS is in units of $1/t$). Since there are no single-electron states with $|\xi_n|>2$, the pair potential does not change as $\hbar\omega_D$ increases above $2$~\cite{bai2023}. Therefore, for the BCS critical temperature we adopt $\hbar\omega_D=1.5$ in Fig.~\ref{fig3}(a) and $\hbar\omega_D=2$ in Fig.~\ref{fig3}(b). 

As shown in Fig.~\ref{fig3}(a) with $\hbar\omega_D=1.5$, the critical temperatures $T_{cs},T_{cb}$, and $T_{c,\,{\rm BCS}}$ remain close to zero for $g<1$. They then exhibit an exponential-like increase in the region $ 1 < g < 2.5$, which is replaced by an almost linear growth with $g$ in the domain $2.5 < g < 5$. Notably, for $g>1.5$, we observe $T_{cs}>T_{cb} > T_{c,{\rm BCS}}$. The difference between $T_{cb}$ and $T_{c,{\rm BCS}}$ is significantly smaller than the difference between $T_{cs}$ and $T_{cb}$, indicating that $T_{cb}$ behaves more similar to the BCS prediction compared to $T_{cs}$. 

In Fig.~\ref{fig3}(b) with $\hbar\omega_D=10$, both $T_{cs}$ and $T_{cb}$ are higher than those shown in Fig.~\ref{fig3}(a) with $\hbar\omega_D=1.5$ for the corresponding coupling values. Although $T_{cs}$ is still larger than $T_{cb}$ in Fig.~\ref{fig3}(b), the gap between them is much less than in panel (a). This indicates that a higher $T_{cs}$ does not necessarily result in a greater surface SC enhancement parameter $\tau$. The underlying physics arises from the complex interplay of higher- and lower-energy quasiparticles, which is sensitive to the Debye energy~\cite{bai2023,bai2023a}. Moreover, both $T_{cs}$ and $T_{cb}$ deviate significantly from the BCS behavior for $g>1.5$, as they are much larger than $T_{c,{\rm BCS}}$. 

\begin{figure}[t]
\centering
\includegraphics[width=1\linewidth]{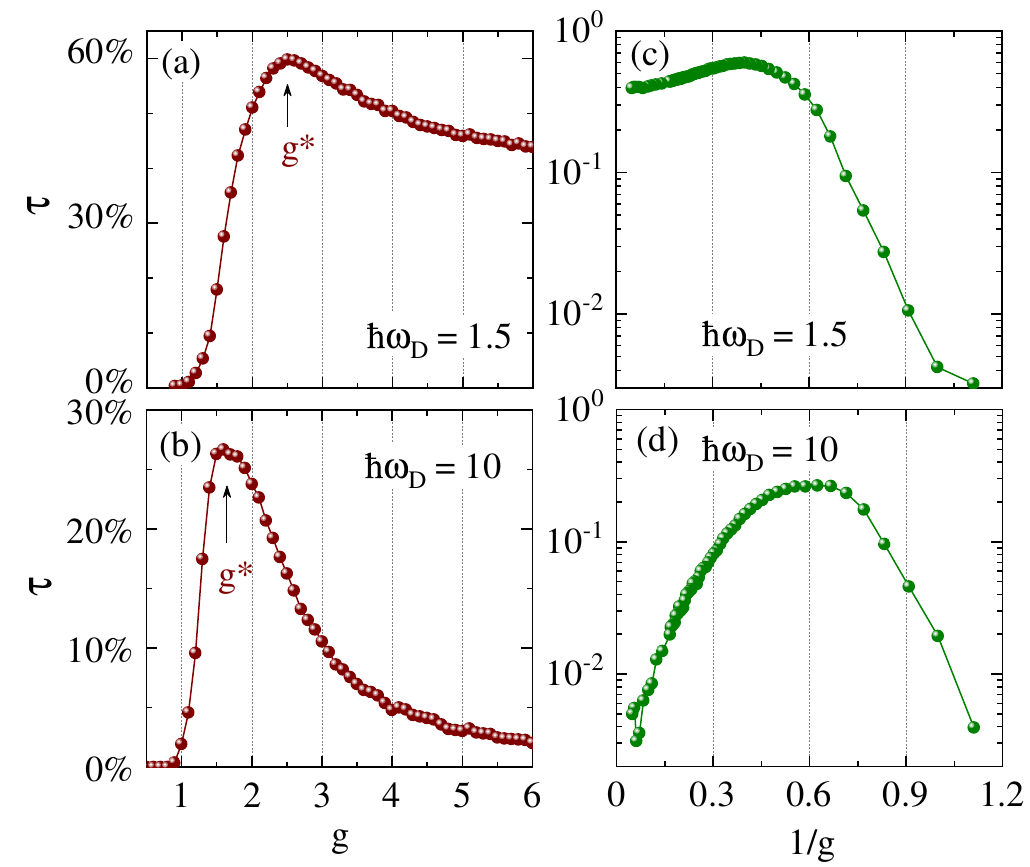}
\caption{The surface SC enhancement parameter $\tau \tau=(T_{cs}-T_{cb})/T_{cb}$ is presented as a function of the coupling strength $g$~(a, b) and $1/g$~(c, d) for $\hbar\omega_D=1.5$ and $10$. Panels (c, d) employ a base-10 logarithmic scale to plot the data for $\tau$. The location of the $\tau$-maximum $g=g*$ can be considered as the crossover point that distinguishes between the weak-coupling and strong-coupling behaviors of $\tau$ as a function of $g$.}
\label{fig4}
\end{figure}

Now, we examine how the surface-SC enhancement parameter $\tau$ depends on the coupling strength $g$ for $\hbar\omega_D=1.5$ and $10$, as shown in Fig.~\ref{fig4}. Panels (a, b) present $\tau$ as a function of $g$ for $\hbar\omega_D=1.5$ and $10$, respectively, while panels (c, d) use a base-10 logarithmic scale to plot the same data as a function of $1/g$, similar to how results for the BCS-BEC crossover are often displayed. For $\hbar\omega_D=10$ in Fig.~\ref{fig4}(b, d), $\tau$ is zero at $g<1$, peaks at $26.7\%$ when $g=g^*=1.60$, and then decreases to zero as $g$ increases beyond $g=g^*$. This qualitatively agrees with the results of Ref.~\cite{samoilenka2020}, which shows that $\tau$ tends to zero in the strong-coupling regime for $\hbar\omega_D > 2$~(all the physical quasiparticles are included in the pair potential) and for various values of the chemical potential. 

In contrast, for $\hbar\omega_D=1.5$, one observes that $\tau$ is exponentially small as $g \to 0$, but increases significantly, reaching its maximum  $\tau_{\rm max}=60.0\%$ when $g$ approaches $g^*=2.5$. For $g>g^*$, $\tau$ gradually decreases as $g$ increases but remains finite, approaching a saturation near $39.4\%$. This differs from the results of Ref.~\cite{samoilenka2020},  highlighting the importance of investigating how the Debye energy influences $\tau$. Notably, the shape of the data in panels (a, c) resembles the behavior of the critical temperature in the BCS-BEC crossover~\cite{perali2004, nozieres1985}. 

\begin{figure}[t]
\centering
\includegraphics[width=0.8\linewidth]{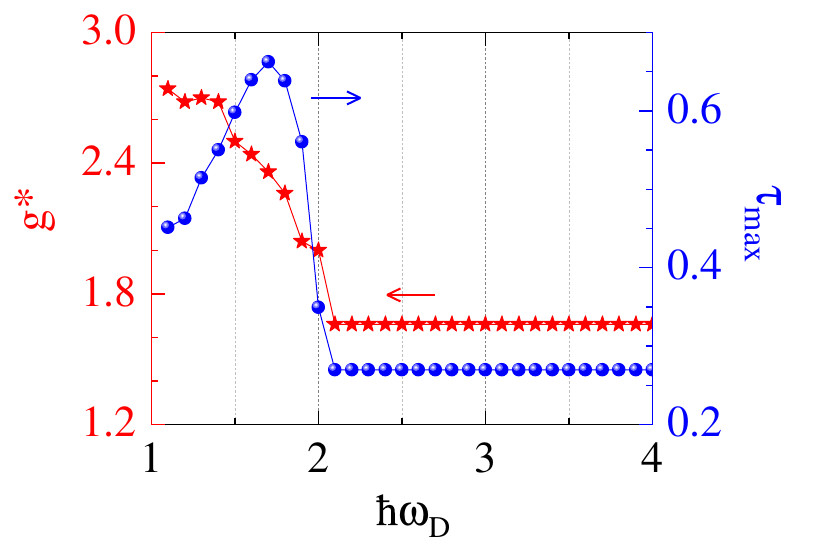}
\caption{The $\hbar\omega_D$-dependence of the characteristics of the universal crossover in the surface SC, with $g^*$~(curves with red stars) and $\tau_{\rm max}$~(curves with blue spheres) plotted as functions of $\hbar\omega_D$.}
\label{fig5}
\end{figure}

Figure~\ref{fig4} demonstrates a crossover in the $g$-dependence of the surface-SC enhancement parameter $\tau$ between qualitatively different weak- and strong-coupling regimes. This crossover is marked by a maximum of $\tau$, denoted as $\tau_{\rm max}$, which occurs at $g=g^*$. Although the crossover is universal and occurs at any value of $\hbar\omega_D$, its characteristics are strongly influenced by a value of the Debye energy. In particular, Fig.~\ref{fig5} demonstrates $g^*$ and $\tau_{\rm max}$ as functions of $\hbar\omega_D$, which are given by red-starred and blue-sphere curves, respectively. One sees, that as $\hbar\omega_D$ increases from $1$ to $2$, $g^*$ decreases, shifting toward the weak-coupling regime. For $\hbar\omega_D > 2$, the position of the maximum of $\tau$ remains unchanged with $\hbar\omega_D$, stabilizing at $g^* = 1.66$. In turn, $\tau_{\rm max}$ initially increases with $\hbar\omega_D$, reaching $66.3\%$ at $\hbar\omega_D = 1.7$, and then it drops rapidly to $26.9\%$, remaining unchanged for $\hbar\omega_D >2$. The peaked behavior of $\tau_{\rm max}$ in Fig.~\ref{fig5} appears because the most significant surface-SC contribution comes from the states with the quasiparticle energy $\varepsilon_\alpha = 1.7$.

\begin{figure}[t]
\centering
\includegraphics[width=1\linewidth]{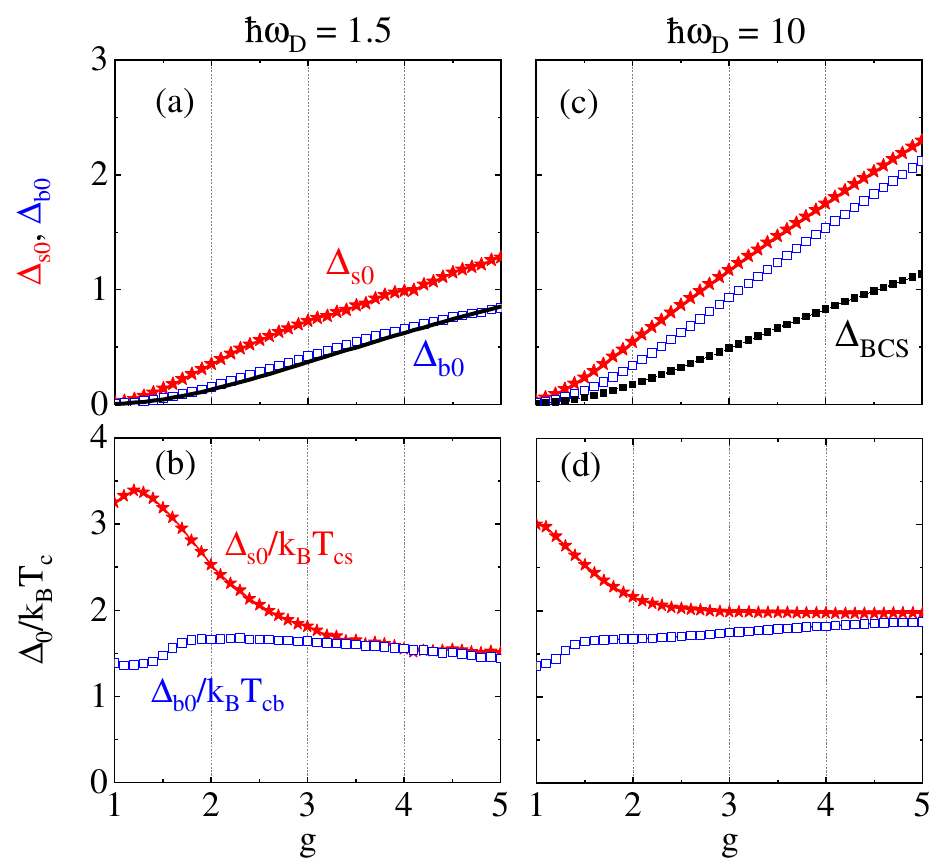}
\caption{$\Delta_{s0}$ and $\Delta_{b0}$~(a, c), and the ratios $\Delta_{s0}/k_BT_{cs}$ and $\Delta_{b0}/k_BT_{cb}$~(b, d) as functions of the coupling strength $g$ for $\hbar\omega_D=1.5$~(a, b) and $10$~(c, d). The surface (for the chain ends) and bulk (for the chain center) results are given by curves with red stars and blue squares, respectively. The BCS estimates of the zero-temperature pair potential $\Delta_{\rm BCS}$ are given by black curves with solid squares, see panels (a) and (c).}
\label{fig6}
\end{figure}

Next, we focus on the $g$-dependence of the ratios $\Delta_{s0}/k_BT_{cs}$, where $\Delta_{s0}=\Delta(i=1)$ is the zero-temperature surface (boundary) pair potential, and $\Delta_{b0}/k_BT_{cb}$, where $\Delta_{b0}=\Delta(i=151)$ denotes the zero-temperature pair potential in the chain center (bulk). In Fig.~\ref{fig6}, the red-starred and blue-squared curves represent $\Delta_{s0}$ and $\Delta_{b0}$, respectively, as functions of $g$ for $\hbar\omega_D=1.5$~(a, b) and $10$~(c, d). Additionally, the black-squared curve corresponds to the BCS result $\Delta_{\rm BCS}$, calculated as $\Delta_{\rm BCS}= 2\hbar\omega_De^{-1/gD(\mu)}$. Following the calculations for $T_{c,{\rm BCS}}$ in Fig.~\ref{fig3}, $\hbar\omega_D$ is taken as 1.5 for $\Delta_{\rm BCS}$ in Fig.~\ref{fig6}(a), while in Fig.~\ref{fig6}(c), $\Delta_{\rm BCS}$ is calculated with $\hbar\omega_D=2$, as $\hbar\omega_D=10$ exceeds half the bandwidth.

From Fig.~\ref{fig6}(a), we observe that for $\hbar\omega_D=1.5$, the zero-temperature pair potential in the chain center, $\Delta_{b0}$, is nearly equal to the BCS estimate $\Delta_{0,{\rm BCS}}$. In contrast, the zero-temperature pair potential at the chain ends, $\Delta_{s0}$, is significantly higher than $\Delta_{0,{\rm BCS}}$. However, for $\hbar\omega_D=10$, which exceeds half the single-electron bandwidth, the difference from the BCS estimate becomes more pronounced: both $\Delta_{s0}$ and $\Delta_{b0}$ are much larger than $\Delta_{0,{\rm BCS}}$. 

Using the results from Figs.~\ref{fig3} and \ref{fig6}(a), we calculate $\Delta_{s0}/k_BT_{cs}$ and $\Delta_{b0}/ k_BT_{cb}$ for $\hbar\omega_D = 1.5$ and $10$, see Figs.\ref{fig6}~(b) and (d). For $\hbar\omega_D = 1.5$, Fig.~\ref{fig6}(b) shows that $\Delta_{s0}/k_BT_{cs}$ increases with $g$ in the region $1 < g < 1.2$, reaching a maximum value of $3.39$ at $g = 1.2$. As $g$ increases from $1.2$ to $3.5$, $\Delta_{s0}/k_BT_{cs}$ decreases rapidly, approaching at $g = 3.5$ the center-of-chain ratio $\Delta_{b0}/k_BT_{cb} = 1.646$, which is lower than the BCS value of about $1.76$~\cite{tinkham1996}. For $g > 3.5$, both $\Delta_{s0}/k_BT_{cs}$ and $\Delta_{b0}/k_BT_{cb}$ exhibit a very slow, almost negligible decrease with increasing $g$. 

The results for $\hbar\omega_D = 10$, shown in Fig.~\ref{fig6}(d), are similar, with two minor differences. First, the maximum value of $\Delta_{s0}/k_BT_{cs}$ is around $3.00$, which is lower than in panel (b). Second, both $\Delta_{s0}/k_BT_{cs}$ and $\Delta_{b0}/k_BT_{cb}$ slowly increase with $g$ for sufficiently strong couplings, approaching $1.92$, which is higher than the corresponding value for $\hbar\omega_D = 1.5$ and the textbook BCS result. Therefore, while $\Delta_{s0}/k_BT_{cs}$ significantly exceeds $\Delta_{b0}/k_BT_{cb}$ on the weak-coupling side of the crossover, they are very close to one another on the strong-coupling side. Notably, $\Delta_{s0}/k_BT_{cs}$ is always larger than the BCS ratio, and significantly exceeds it in the weak-coupling regime.

\section{Conclusions}~\label{IV}

By numerically solving the self-consistent Bogoliubov-de Gennes equations for a $1D$ attractive Hubbard model with $s$-wave pairing at half-filling, we have systematically investigated the combined effects of varying both the pair coupling $g$ and the Debye energy $\hbar\omega_D$ on surface superconductivity (SC) enhancement. Our study reveals a universal crossover in surface SC from weak- to strong-coupling behavior, which occurs regardless of the specific value of $\hbar\omega_D$. This crossover is characterized by a maximum in the relative enhancement of the surface critical temperature, $\tau$ (the critical temperature at the chain end in our study), which can reach up to about $70\%$.

The characteristics of this crossover are highly sensitive to the Debye energy when it is below half the single-particle bandwidth, with the enhancement of surface SC being most pronounced in this case. While our study focuses on half-filling, similar results are expected beyond this condition, although the effect tends to diminish as the system deviates from half-filling~\cite{croitoru2020,samoilenka2020a}.

Although our investigation is based on a $1D$ model, a similar universal crossover can be expected at any value of $\hbar\omega_D$ for $2D$ and $3D$ attractive Hubbard models, as can also be seen from conclusions of Ref.~\onlinecite{samoilenka2020a}. In particular, the results of our model are applicable to the case of a Hubbard model describing an array of parallel chains with relatively small hopping amplitudes between them, see Ref.~\cite{wang2024b}.  Moreover, this kind of universal crossover in the coupling dependence of the critical temperature could also be expected for corner SC enhancements in $2D$ and $3D$ cases, also reported for an attractive Hubbard model~\cite{samoilenka2020a}.



\begin{acknowledgments}
This work was supported by the Science Foundation of Zhejiang Sci-Tech University (Grants No. 19062463-Y). A.A.S. thanks the Grant of Ministry of Science and Higher Education of the Russian Federation No. 075-15-2024-632 for the support that helped to perform investigations of the universal crossover in the interference-induced surface superconductivity. Analysis of numerical solutions of the formalism was supported by the HSE University Basic Research Program.
\end{acknowledgments}


\end{document}